\batchmode
\makeatletter
\def\input@path{{C:/Users/idanl/Documents/Thesis/}}
\makeatother
\documentclass[twocolumn,english]{revtex4-2}
\usepackage[T1]{fontenc}
\usepackage[latin9]{inputenc}
\usepackage{refstyle}
\usepackage{amsmath}
\usepackage{amssymb}
\usepackage{graphicx}
\usepackage{wasysym}

\makeatletter

\AtBeginDocument{\providecommand\tabref[1]{\ref{tab:#1}}}
\providecommand{\tabularnewline}{\\}
\RS@ifundefined{subsecref}
  {\newref{subsec}{name = \RSsectxt}}
  {}
\RS@ifundefined{thmref}
  {\def\RSthmtxt{theorem~}\newref{thm}{name = \RSthmtxt}}
  {}
\RS@ifundefined{lemref}
  {\def\RSlemtxt{lemma~}\newref{lem}{name = \RSlemtxt}}
  {}

\usepackage{tikz}
\usetikzlibrary{quantikz2}
\usepackage{titlesec}
\titlespacing\paragraph{8pt}{0pt plus 2pt minus 2pt}{3pt plus 1pt minus 1pt}

\makeatother

\usepackage{babel}
\begin{document}
\title{Ensemble Dependence of the Critical Exponent at a Quantum Error Correction
Threshold}
\author{Idan Dror and Moshe Goldstein}
\affiliation{School of Physics and Astronomy, Tel-Aviv University, Tel Aviv 6997801,
Israel}
\begin{abstract}
In thermodynamics it is common to assume that the choice of ensemble
(e.g., micro-canonical, canonical, or grand-canonical) should not
affect the underlying physics in the thermodynamics limit. We show
that this does not necessarily hold for critical exponents. Examining
a simplified model of quantum error correction (single step encoding
and decoding by a random unitary) and the behavior of both the fidelity
and magic at the corresponding threshold, we find different exponents
when using generic channels or supposedly equivalent quantum trajectories.
Interestingly, the obtained exponents saturate a recently derived
information theoretic bound \citep{feldman2024informationboundsphasetransitions}
which we extend from the grand-canonical to the canonical case, including
intermediate ensembles which we define. Moreover, even the existence
of the transition is shown to be ensemble-dependent.
\end{abstract}
\maketitle
The universality properties of phase transitions \citep{RevModPhys.46.597,DePalma2025,PhysRevE.56.2310}
across a wide range of thermodynamic systems has long been a fruitful
principle in many-body physics. One of the most notable implications
of universality is that critical exponents are mostly indifferent
to the details of the system \citep{10.1143/PTP.51.1992,RevModPhys.71.S358}.
Since it is well known that, in the thermodynamic limit, different
thermodynamic ensembles (micro-canonical, canonical, and grand-canonical)
provide the same predictions for physical quantities (unless they
are directly affected by the ensemble, that is, vanish in one but
not the other \citep{Behringer_2005}), it is only natural to assume
that critical exponents (of unconstrained quantities) are universal
with respect to the choice ensemble as well, at least when the interactions
are short range (long range systems are known to be exceptions \citep{PhysRevLett.133.050403,Campa_2025,PhysRevLett.134.207401,Vodola_2022}).

The situation turns out to be more subtle when discussing ensembles
with respect to impurities (e.g., fixed vs. fluctuating number of
impurities). When disorder is annealed, that is, when its distribution
is of the Boltzmann form, the usual proof of ensemble equivalence
should hold, though see Ref. \citep{1968PhRv..176..257F}. When disorder
is quenched the usual proof does not hold, but one would still expect
that ensemble difference will be negligible when the number of impurities
is extensive; a possible exception \citep{Dhar_2003} was shown to
pertain to only edge and not bulk observables, and to result from
vanishing leading order coefficient for the former \citep{Igloi_1998}.

Recent years have seen a surge of interest in quantum computation,
which is expected to yield significant advantages over classical computation
in many applications \citep{Preskill2018quantumcomputingin}, while
providing a perhaps rich view perspective of modern many-body physics.
Quantum error correction \citep{PhysRevA.52.R2493} is vital to the
field due to its role in lowering the strict requirements on the performance
of quantum computers needed to achieve fault-tolerant computation
\citep{shor1997faulttolerantquantumcomputation,doi:10.1137/S0097539799359385}.
A n important aspect of both classical and quantum error correction
is the error-correction threshold, which is the critical error strength
(usually given in terms of an error probability), above which an error
correction scheme degrades the probability of successful decoding
instead of improving it. As such, it behaves much like a phase transition,
an observation has been made by various previous works \citep{PhysRevLett.132.140401,liu2026coherenterrorinducedphase,PhysRevX.9.031009,PhysRevLett.125.030505,nelson2025errorcorrectionphasetransition}.
A common system of interest for quantum computation and many-body
physics in general and quantum error correction in particular is random
unitary with errors \citep{PhysRevB.111.064308}, where the time evolution
of a quantum register is given by random unitary operations, commonly
taken as random two-qudit gates in a brick-wall like structure, interleaved
with various types of errors, most notably non-unitary errors such
as projective measurements. The study of such systems is aided by
thermodynamic approaches \citep{Vodola_2022}, since each instance
of the circuit can be intractable to simulate, but the statistics
of the system can be well understood and provide insight to many body
dynamics, such as the phase transition at the quantum error-correction
threshold. Non-unitary errors are of a particular interest due to
being a more general error type that is likely to occur in practical
applications, such as entanglement with an environment and qudit loss.

In this paper, we show that there are cases where the scaling of a
quantity can depend strongly on the ensemble despite common belief
and under less restrictive conditions than previous papers, through
an ensemble dependence of a critical exponent. Specifically, we demonstrated
that a simplified quantum error-correction model exhibits a phase
transition at the error-correction threshold, such that its success
rate, as well as the magic of the resulting state, depends on a parameter
which scales as $N^{1/\nu}$, where $N$ is the number of qudits,
and the critical exponent $\nu$ takes the value $\nu=1$ for the
canonical ensemble (similarly to the deterministic case considered
before \citep{PhysRevLett.132.140401,sierant2026theorymagicphasetransitions}),
and $\nu=2$ for the grand-canonical ensemble. A summary table of
the critical exponent bounds and exact value in different ensembles
is presented in \tabref{The-critical-exponent}. 

\begin{table}[h]
\begin{tabular}{|c|c|c|}
\hline 
Ensemble &
Bound &
Exact value\tabularnewline
\hline 
\hline 
canonical/deterministic &
$\nu\geq1$ &
$\nu=1$\tabularnewline
\hline 
grand-canonical &
$\nu\geq2$ &
$\nu=2$\tabularnewline
\hline 
\end{tabular}

\caption{The critical exponent information-theoretic bound and exact value
for different ensembles assuming quenched averaging over the error
distribution.\label{tab:The-critical-exponent}}
\end{table}

We note that ensemble difference has been conjectured in a closely
related system before, based on numerical simulations or analytical
assumptions \citep{d4wh-hqcp}; see also \citep{PhysRevLett.79.5130}.
Here we prove the result rigorously. We also show that one may define
ensembles interpolating between canonical and grand-canonical with
continuously varying bound on $\nu$. Moreover, we find that even
the existence of the transition may depend on the ensemble.

\begin{figure}
\begin{quantikz}
\lstick{\ket{\psi}} & \qwbundle{k}   & \gate[2]{U} && \gate[2,style=circle]{\mathcal{E}} && \gate[2]{U^{\dagger}} & \\
\lstick{\ket{0}}    & \qwbundle{N-k} &             &&                                    && &
\end{quantikz}

\caption{\label{system diagram} Schematic diagram of the system, comprised
of $N$ qudits, the first $k$ of which are logical and the other
are ancillary. The state is encoded using a unitary $U$, then a global
channel $\mathcal{E}$ is applied, and finally the state is decoded
by the inverse $U^{\dagger}$.}
\end{figure}
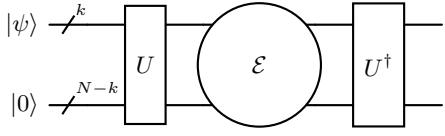

Our model is a natural extension of the encoding-decoding circuit
described in \citep{PhysRevLett.132.140401}. Given some logical state
$\left|\psi\right\rangle $ in a $K$ dimensional Hilbert space, that
state is encoded into a $D>K$ space by applying a random unitary
$U\in\text{U}\left(D\right)$ on the state, extended by some ancilla,
which would hold error syndromes upon decoding. Symbolically, $\left|\psi\right\rangle \mapsto U\left(\left|\psi\right\rangle \left|0\right\rangle \right)$.
Without loss of generality, we take $\left|\psi\right\rangle =\left|0\right\rangle $.
A random error, represented as a quantum channel $\mathcal{E}_{s}$,
is applied to the system with probability $p_{s}$, where the index
$s$ enumerates all of the errors in the selected distribution. Decoding
is then attempted by applying the inverse unitary $U^{\dagger}$.
A schematic diagram of the system is presented in Fig. \ref{system diagram}.
The final state of the system $\rho$ is therefore takes the form
$U^{\dagger}\mathcal{E}_{s}\left(U\left|0,\bar{0}\right\rangle \left\langle 0,\bar{0}\right|U^{\dagger}\right)U$
with probability $p_{s}$, and the bar denotes the ancillary subsystem.
If the ancilla return to their initial state, no syndrome arises,
and no error correction is applied. In this situation, the fidelity
of the final state with initial state (projected into the zero-ancilla
subspace), is given by

\begin{equation}
F\equiv\mathbb{P}\left(0\cap\bar{0}\middle|\bar{0}\right)=\frac{\left\langle 0,\bar{0}\middle|\rho\middle|0,\bar{0}\right\rangle }{\text{Tr}\left\langle \bar{0}\middle|\rho\middle|\bar{0}\right\rangle }.\label{eq:fidelity_def}
\end{equation}
Assuming similar fidelity if syndromes are on and error correction
is activated, Eq. (\ref{eq:fidelity_def}) can be used as a proxy
for the probability of a successful error-correction. As in \citep{PhysRevLett.132.140401},
we average the fidelity over the probability distribution of $U$,
which is assumed to be the Haar measure. In addition, whereas Ref.
\citep{PhysRevLett.132.140401} considered (for the most part) a deterministic
error channel, we average the fidelity over the error distribution
$p_{s}$. The result is a single value for the fidelity for each distribution
of errors. 

In particular, we focus on the case where the system is composed of
$N$ sites (representing qudits of dimension $d$), $k$ of these
$N$ qudits are logical, so that dimension of the entire space is
$D=d^{N}$ and of the logical subspace is $K=d^{k}$. Moreover, we
assume that the possible errors channels are separable between the
sites $\mathcal{E}_{\vec{\mu}}=\bigotimes_{i=1}^{N}\mathcal{E}_{\mu_{i}}$.
The thermodynamic limit is therefore $N\to\infty$, while keeping
the encoding rate $r\equiv\frac{k}{N}$ constant. In the limit, the
fidelity exhibits a phase transition between $F=1$ for ``weak''
errors to $F=0$ for ``strong'' errors, with a scaling behavior,
$F=F\left(N^{1/\nu}\left(\lambda-\lambda_{c}\right)\right)$, where
$\lambda$ is the disorder parameter, which is an arbitrary notion
of error strength, $\lambda_{c}$ is the critical value, and $\nu$
is the critical exponent, which is of particular interest for us,
as we will show that it has different values for different ensembles,
and that an ensemble average might even eliminate the transition altogether,
shifting $\lambda_{c}$ to zero. We note that the global random unitary
can be replaced by a brick-wall circuit of local two-qudit unitaries,
which we employ in our simulations below, hence the system is effectively
local and short-ranged.

Another quantity of interest is a measure of magic, the stabilizer
R nyi entropy (SRE), defined by Ref. \citep{PhysRevLett.128.050402}
as

\begin{equation}
M_{q}=\frac{1}{1-q}\sum_{P\in\mathcal{P}_{k}}\frac{\text{Tr}\left[\left(\rho_{L}P\right)^{2q}\right]}{2^{k}},
\end{equation}
where $\mathcal{P}_{k}$ denotes the group of Pauli strings on $k$
qubits, and $\rho_{L}=\frac{\left\langle \bar{0}\middle|\rho\middle|\bar{0}\right\rangle }{\text{Tr}\left\langle \bar{0}\middle|\rho\middle|\bar{0}\right\rangle }$
is the logical part of the density matrix, post-selected to vanishing
ancillae. We show this quantity to exhibit the same ensemble-dependent
exponent $\nu$ as the fidelity.

\paragraph{Taking averages.---}

The most natural way to average over random distributions (of the
encoding unitary $U$ and the error channel $\alpha$), the so-called
``quenched'' average is

\begin{equation}
F_{2}\equiv\mathbb{E}_{s,U}\left[\frac{\mathbb{P}\left(0\cap\bar{0}\right)}{\mathbb{P}\left(\bar{0}\right)}\right],\label{F_2 quenched}
\end{equation}
however, it is hard to work with analytically. To simplify, we can
average over the encodings separately for the numerator and denominator,

\begin{equation}
\tilde{F}_{2}=\mathbb{E}_{s}\left[\frac{\mathbb{E}_{U}\left[\mathbb{P}\left(0\cap\bar{0}\right)\right]}{\mathbb{E}_{U}\left[\mathbb{P}\left(\bar{0}\right)\right]}\right].
\end{equation}
 This simpler version is called the ``annealed'' (with respect to
$U$) fidelity, and a method for calculating it is described in \citep{PhysRevLett.132.140401,sierant2026theorymagicphasetransitions},
which also showed that both agree in the thermodynamic limit for deterministic
errors, and hence we expected them to agree for random errors as well,
meaning that $F_{2}$ is self-averaging.

We can also apply the quenched/annealed dichotomy to the error types:

\begin{equation}
F_{1}=\mathbb{E}_{U}\left[\frac{\mathbb{E}_{s}\left[\mathbb{P}\left(0\cap\bar{0}\right)\right]}{\mathbb{E}_{s}\left[\mathbb{P}\left(\bar{0}\right)\right]}\right],\label{F_1 quenched}
\end{equation}

\begin{equation}
\tilde{F}_{1}=\frac{\mathbb{E}_{s,U}\left[\mathbb{P}\left(0\cap\bar{0}\right)\right]}{\mathbb{E}_{s,U}\left[\mathbb{P}\left(\bar{0}\right)\right]}.\label{F_1 annealed}
\end{equation}
However, they are simply equivalent to a single deterministic error.
Given an initial state $\rho$ and a distribution of channels, the
probability of measuring a state $x$ is $\mathbb{P}\left(x\right)=\sum_{s}p_{s}\left\langle x\middle|\mathcal{E}_{s}\left(\rho\right)\middle|x\right\rangle =\left\langle x\middle|\sum_{s}p_{s}\mathcal{E}_{s}\left(\rho\right)\middle|x\right\rangle $,
which is the same as the probability given a single deterministic
``mixture'' channel $\mathcal{E}_{\text{mix}}\left(\rho\right)\equiv\sum_{s}p_{s}\mathcal{E}_{s}\left(\rho\right)$.
As we will see below, proper quenched averaging over $s$ could lead
to different results, including the absence of a transition or a change
of the critical exponents.

\paragraph*{Choice of ensemble.---}

There are many possible distributions of the random channels, but
two of particular interest are those corresponding to the classical
thermodynamic ensembles, specifically the canonical and grand-canonical
ensembles. In the grand-canonical ensemble, the randomization of errors
is done in a site-independent manner, such that the total number of
errors of each type is allowed to vary. We denote the single-qudit
error channels $\mathcal{E}_{\alpha}$, and the probability of an
arbitrary site to have error of type $\alpha$ as $p_{\alpha}$. Conversely,
in the canonical ensemble, the number of errors of each type is predetermined,
and randomness can only affect the order of errors. The total number
of errors of type $\alpha$ is set to be the expected number of errors
of that type in the grand-canonical ensemble, i.e. $p_{\alpha}N$.

\paragraph{Bounding the critical exponent.---}

It is possible to bound the critical exponent without having to explicitly
calculate it using the information-theoretic approach described in
\citep{feldman2024informationboundsphasetransitions}. Given a large
enough system, a simulation could be used to determine its phase.
In order to initialize the simulation, a specific number of samples
is needed from the error distribution. Let us first determine the
minimal number of samples needed to distinguish the phases. Assuming
that close to the transition point, the expected fidelity takes the
form $\bar{F}=f\left(N^{1/\nu}\delta\right)$, where $\delta\equiv\lambda-\lambda_{c}$,
and where the universal critical exponent $\nu$ and the universal
scaling function $f$ are assumed to be known, with $f$ being monotonic
near the critical point. For any value of $\left|\delta\right|>0$
we take $N\sim\delta^{-\nu}$, giving us just enough sensitivity to
distinguish between $f\left(\pm1\right)$. Numerically estimating
that value with a precision of order $\epsilon\apprle\frac{1}{2}\min\left|f\left(\pm1\right)-f\left(0\right)\right|$
tells us the phase. According to Chebyshev's inequality, $\mathbb{P}\left(\left|F-\bar{F}\right|>\epsilon\right)\leq\frac{\text{Var}\left(F\right)}{\epsilon^{2}}$.
Note that by using the expected fidelity, we restrict ourselves to
the quenched-error variants $F_{2}$, $\tilde{F}_{2}$. If we assume
that $\lim_{N\to\infty}\text{Var}\left(F\right)=0$, then there exists
$\delta>0$ and a corresponding $N\sim\delta^{-\nu}$ such that $\text{Var}\left(F\right)\apprle\epsilon^{2}$.
For such small $\delta$, the probability of getting the correct phase
from a single simulation is $O\left(1\right)$, and therefore a finite
amount of repetitions is enough to know the phase with any required
confidence. Assuming that the error channel on each site is independent
of the others, one sample is required for each site, meaning that
the number of samples is $N_{\text{PT}}\propto N\sim\delta^{-\nu}$.

Using the phase transition to try and find the sign of $\delta$ is
inherently wasteful, as the data is distorted by $f$. The most efficient
way to extract the sign of $\delta$ is to use the samples from the
distribution directly. Such a strategy can be shown to need at least
$N_{\text{opt}}\sim\delta^{-2}$ samples using elementary statistics.
Using the fact that $N_{\text{PT}}\geq N_{\text{opt}}$, we deduce
a bound for the critical exponent is $\nu\geq2$, which is the Harris
criterion \citep{A_B_Harris_1974,PhysRevLett.57.2999,Chayes1989}
for a one dimensional system.

The assumption of independent errors fixes the ensemble to be the
grand-canonical. In the average over the encoding space, we can always
swap the order of sites by applying a unitary transformation, to which
the Haar measure is invariant. Since the order of the sites determines
the order of the errors, the order of the errors cannot affect the
encoding-average fidelity, so it must depend only on the number of
errors of each type. As $N\to\infty$, the variances of these numbers
vanish, and therefore we expect $\text{Var\ensuremath{\left(F\right)}\ensuremath{\ensuremath{\to}}}0$,
which is the other required assumption. Based on the above arguments,
we expect $\nu\geq2$ for $F_{2},\tilde{F}_{2}$ when the errors are
taken from the grand-canonical ensemble. As we will show below, in
fact this bound not only holds, but it is also saturated for $\tilde{F}_{2}$,
meaning $\nu=2$. This is similar to the case of making the channel
parameter an independent random variable for each qubit, which was
found in Ref. \citep{PhysRevLett.132.140401} to give rise to $\nu=2$.
the error-annealed $F_{1},\tilde{F}_{1}$, Chebyshev's bound does
not hold, as the fidelity is not the mean of the distribution.

To deal with the cases for which the above method does not apply,
we propose another similar yet distinct argument, which yields a looser
bound but is independent of the ensemble and averaging method. This
bound turns out to be saturated by $\tilde{F}_{1}$ and the canonical
ensemble, as observed in Ref. \citep{PhysRevLett.132.140401}. Let
$\tilde{p}\left(\lambda,M\right)$ be the expected proportion of a
specific outcome (in our case, a specific type of error) of the distribution
$\mathcal{P}_{\lambda}$ when taking $M$ (not necessarily independent)
samples, and let $p\left(\lambda\right)=\lim_{M\to\infty}p\left(\tilde{\lambda},M\right)$
the same proportion in the the thermodynamic limit. Since $M$ is
by definition an integer, the accuracy to which $p$ can be estimated
by sampling the distribution is given by $\delta\sim M^{-1}$. Therefore,
to measure $p$ within $\delta$ of its true value, one must take
$M_{\text{opt}}\apprge\delta^{-1}$ samples, regardless of the distribution.
On the other hand, obtaining an accuracy of $\delta$ from a phase
transition simulation, through similar reasoning as the grand-canonical
ensemble, still requires $M_{\text{PT}}=N\apprge\delta^{-\nu}$ samples
near the phase transition. Since the phase transition cannot be more
accurate than probing the distribution directly, $M_{\text{PT}}\geq M_{\text{opt}}$,
and thus $\nu\geq1$.

It is possible to extend the above arguments to encompass a continuum
of ensembles between the canonical ensemble, which has a fully dependent
distribution, and the grand-canonical ensemble, which has fully independent
distribution. When taking a reservoir of errors $N_{\text{bath}}$
from which $N$ are sampled without return, and $N_{\text{hidden}}=N_{\text{bath}}-N\sim N^{a}$
are therefore left hidden for some $a\geq0$, it turns out the the
critical exponent obeys 
\begin{equation}
\nu\geq\frac{2}{2-\min\left(a,1\right)}.\label{eq:continuum-ensemble}
\end{equation}
It thus interpolates between a pseudo-canonical ensemble phase for
$a=0$ (constant $N_{\text{hidden}}$), and a pseudo grand-canonical
ensemble for $a\geq1$. The derivation of this bound is in \citep{sm}.

While the above arguments referred to the fidelity, the same reasoning
applies to other quantities, provided that the satisfy the appropriate
condition for each type of bound. This includes the R nyi entropy
and the SRE.

\paragraph{Analytic calculation of the fidelity.---}

To calculate the encoding-annealed fidelity, we employ the method
described by \citep{PhysRevLett.132.140401}. For a general deterministic
error, this yields (see \citep{sm}):

\begin{equation}
\tilde{F}\left(\mathcal{E}\right)=\frac{\left(D-1\right)\left(DR+1\right)}{DR\left(D-D^{r}\right)+D^{1+r}-1}\to_{D\gg1}\frac{1}{1+\frac{D^{r-1}}{R}},\label{annealed F exact}
\end{equation}
 where we defined $R$ as the ``average similarity'' of a state
before and after applying the channel $R$:

\begin{equation}
R\left(\mathcal{E}\right)=\frac{1}{D^{2}}\sum_{i,j=1}^{D}\left\langle i\right|\mathcal{E}\left(\left|i\right\rangle \left\langle j\right|\right)\left|j\right\rangle ,
\end{equation}
 and the indices go over an orthonormal basis of the system's Hilbert
space. This quantity has the following key properties:
\begin{itemize}
\item Independence of the choice of Kraus operators and choice of basis,
making it a property of the channel itself.
\item Bound, $0\leq R\leq1$, with the identity channel being the only channel
to have $R=1$, while only channels with only traceless Kraus operators
have $R=0$.
\item Multiplicative under the Kronecker product, $R\left(\otimes_{i}\mathcal{E}_{i}\right)=\prod_{i}R\left(\mathcal{E}_{i}\right)$.
Note that the RHS is order independent, due to the qudit-swapping
symmetry of encoding-average setup.
\item Linearity, $R\left(\sum_{\alpha}p_{\alpha}\mathcal{E}_{\alpha}\right)=\sum_{\alpha}p_{\alpha}R\left(\mathcal{E}_{\alpha}\right)$.
\end{itemize}
As stated above, $R\left(\mathcal{E}\right)$ quantifies the ``average''
similarity of the output states to the input states, not how much
of the original information is retained, as unitary channels can have
$R=0$. Since stronger errors lead to different states, and errors
add up, it proves useful to define the error strength as $E\left(\mathcal{E}\right)\equiv-\log_{D}R\left(\mathcal{E}\right)$.
We may then rewrite (\ref{annealed F exact}) in terms of $E$:

\begin{equation}
\tilde{F}\left(\mathcal{E}\right)\approx\frac{1}{1+D^{E-\left(1-r\right)}}.
\end{equation}

For a multi-site system, $D=d^{N}$ and $\mathcal{E}=\bigotimes_{i}\mathcal{E}_{i}$,
a phase transition emerges:

\begin{equation}
\tilde{F}\left(\mathcal{E}\right)\approx\frac{1}{1+d^{N^{1/\nu}\left(\bar{E}-E_{c}\right)}}\to\Theta\left(-\left(\bar{E}-E_{c}\right)\right),\label{F_multi exact}
\end{equation}
where $\bar{E}=\frac{1}{N}\sum_{i}E\left(\mathcal{E}_{i}\right)$
is the average error strength, its critical value being $E_{c}=1-r$,
the ``redundancy rate'', $\nu=1$ is the critical exponent, and
$\Theta$ is the Heaviside step function (defined to have $\Theta\left(0\right)=\frac{1}{2}$).
In this form it is clear how when the error strength exceeds the redundancy
of the system, the error-correcting threshold is broken and error
correction fails. Multiplying both sides by $N$, the phase transition
occurs when the total error reaches the number of ancillae. This is
analogous to the Singleton bound \citep{782103,PhysRevA.55.900,JOSHI1958289,1053661},
with the total error standing in for the number of errors. Note that
$E_{c}<1$, but $\bar{E}$ is unbound, which shows that there are
errors that are too strong to correct no matter how much redundancy
exists in the system when using Haar-random unitary encoding without
recovery operations.

When adding randomness, $\tilde{F}_{1}$ still behaves in a deterministic
manner. The two ensembles differ by the equivalent deterministic channels.
For the canonical ensemble, that is $\mathcal{E}_{\text{canon}}=\bigotimes_{\alpha}\mathcal{E}_{\alpha}^{\otimes\left(p_{\alpha}N\right)}$,
one recovers Eq. (\ref{F_multi exact}) with $\bar{E}=\sum_{\alpha}p_{\alpha}E\left(\mathcal{E_{\alpha}}\right)$.
For the grand-canonical-ensemble, $\mathcal{E_{\text{grand}}}=\left(\sum_{\alpha}p_{\alpha}\mathcal{E_{\alpha}}\right)^{\otimes N}$
, and therefore (\ref{F_multi exact}) still hold, except that the
natural parameter changes to $\tilde{E}=E\left(\sum_{\alpha}p_{\alpha}\mathcal{E_{\alpha}}\right)$,
the error strength of the average single-site channel. In both cases,
the critical exponent is $\nu=1$, the critical value is $E_{c}=1-r$,
and the limiting form is the same, but the choice of ensemble changes
the error strength.

Recalling again that the Haar-randomness of the encoding leads to
permutation symmetry of the single-qudit channels, one gets that in
the canonical ensemble $\tilde{F}_{2}=\tilde{F}_{1}$, and the randomness
is still artificial. However, in the grand-canonical ensemble, the
average error $\bar{E}$ is normally distributed due to the central
limit theorem, resulting in (see \citep{sm}):

\begin{align}
\tilde{F}_{2} & \approx\frac{1}{2}\left(\text{erf}\left(-\sqrt{\frac{N}{2\text{Var}\left(E\right)}}\left(\bar{E}-E_{c}\right)\right)+1\right)\nonumber \\
 & \to\Theta\left(-\left(\bar{E}-E_{c}\right)\right),
\end{align}
which has $\nu=2$ as expected from the derived bound, and the same
critical value $E_{c}=1-r$ and error strength $\bar{E}=\sum_{\alpha}p_{\alpha}E\left(\mathcal{E}_{\alpha}\right)$
as the canonical ensemble. This approximation is only valid when $N\gg\frac{1}{\text{Var}\left(E\right)}$,
as when the variance of the error strengths is small, the result approaches
that of a deterministic channel with the average error strength. Note
that where $\text{Var}\left(E\right)=0$, there is no distinction
between ensembles and error-average type, so the critical exponent
drops to $\nu=1$. 

A concrete example in which one could see both $\nu=1$, $\nu=2$,
and no phase transition, is the depolarization channel $\mathcal{E}_{\lambda}\left(\rho\right)=\lambda\frac{I}{d}+\left(1-\lambda\right)\rho$.
For $d=2$, it can be decomposed as a sum of unitary channels as
\begin{align}
\mathcal{E}_{\lambda}\left(\rho\right) & =\left(1-\frac{3\lambda}{4\sin^{2}\frac{\phi}{2}}\right)\rho\nonumber \\
 & \quad+\frac{1}{2}\sum_{i=x,y,z}\frac{\lambda}{4\sin^{2}\frac{\phi}{2}}e^{+i\frac{\phi}{2}\sigma_{i}}\rho e^{-i\frac{\phi}{2}\sigma_{i}}\nonumber \\
 & \quad+\frac{1}{2}\sum_{i=x,y,z}\frac{\lambda}{4\sin^{2}\frac{\phi}{2}}e^{-i\frac{\phi}{2}\sigma_{i}}\rho e^{+i\frac{\phi}{2}\sigma_{i}},
\end{align}
where $0<\phi\leq\pi$ is an arbitrary parameter. The case of $\phi=\pi$
is a standard representation of the depolarization channel, which
is used in \citep{PhysRevLett.132.140401}. We will select $\mathcal{E}_{\pm i}=e^{\pm i\frac{\phi}{2}\sigma_{i}}$
with $p_{\pm i}=\frac{1}{2}\frac{\lambda}{4\sin^{2}\frac{\phi}{2}}$,
and $\mathcal{E}_{0}=I$ with $p_{0}=1-\frac{3\lambda}{4\sin^{2}\frac{\phi}{2}}$.
Using $R\left(\mathcal{E}_{\pm i}\right)=\cos^{2}\frac{\phi}{2}$
and $R\left(\mathcal{E}_{0}\right)=1$, we find 
\begin{equation}
\bar{E}=-\frac{3\lambda}{4}\frac{\log_{2}\left(\cos^{2}\frac{\phi}{2}\right)}{\sin^{2}\frac{\phi}{2}},
\end{equation}

\begin{equation}
\tilde{E}=-\log_{2}\left(1-\frac{3\lambda}{4}\right),
\end{equation}
which shows explicitly how the fidelities depend on the realization
$\phi$ of the channel through $\bar{E}$, except for the grand-canonical
case of $\tilde{F}_{1}$, which involves $\tilde{E}$ and therefore
does not depend on $\phi$, as expected. When $\phi=\pi$, the average
error-strength is $\bar{E}=\infty$ for $\lambda\neq0$, essentially
eliminating the phase transition, despite originating from the same
mixture channel, as even a single non-identity error is enough to
thwart error-correction completely in this model.

Ref. \citep{sierant2026theorymagicphasetransitions} describes the
same phase transition also in terms of the SRE. Through similar analysis,
they show that for unitary product channel $\mathcal{E}\left(\rho\right)=\left(e^{i\frac{\alpha}{2}\sigma_{z}}\right)^{\otimes N}\rho\left(e^{-i\frac{\alpha}{2}\sigma_{z}}\right)^{\otimes N}$
the SRE of the logical subspace post-selected on zero ancilla in the
thermodynamic limit is approximately

\[
\tilde{M}_{q}\left(\alpha\right)\propto-\log_{2}\left(\frac{1}{1+\frac{2^{N\left(r-1\right)}}{R\left(\alpha\right)}}\right)=-\log_{2}\tilde{F}\left(\alpha\right),
\]
with $R\left(\alpha\right)=\cos^{2N}\left(\frac{\alpha}{2}\right)$.
Generalizing to independent angles, one can get $R\left(\alpha\right)=\prod_{i=1}^{N}\cos^{2}\left(\frac{\alpha_{i}}{2}\right)$
and retrieve $\tilde{M}_{q}^{\left(1,2\right)}\propto-\log_{2}\tilde{F}_{1,2}$.
Therefore, the same analysis applies, leading in particular to the
same jump from $\nu=1$ in the canonical ensemble to $\nu=2$ in the
grand-canonical ensemble.

\paragraph{Numerical results.---}

We present the result of simulating $F_{1}$ and $F_{2}$ (encoding-quenched)
for the distribution associated with the $z$-only depolarization
channel ($\sigma_{i}\mapsto\sigma_{z}$) for $\phi=\frac{\pi}{2}$
as defined in the previous section. This channel has the same error
strength as the full depolarization channel, but allows a more gradual
change in error strength through fewer unitaries in the canonical
ensemble. The sampling of $U$ is done through $2N$ layers of local
random 2-qubit gates, arranged in a brick-wall pattern, to match the
procedure of Ref. \citep{PhysRevLett.132.140401}. Since the encoding-annealed
version have been calculated analytically, we only simulated the quenched
variant. The critical exponents and values were calculated using the
analytical results for the annealed variant, and can be seen to match
the quenched results as expected. The simulation is performed using
state-vector formalism since all possible trajectories are taken to
be unitary. 

\begin{figure}
\includegraphics[scale=0.28]{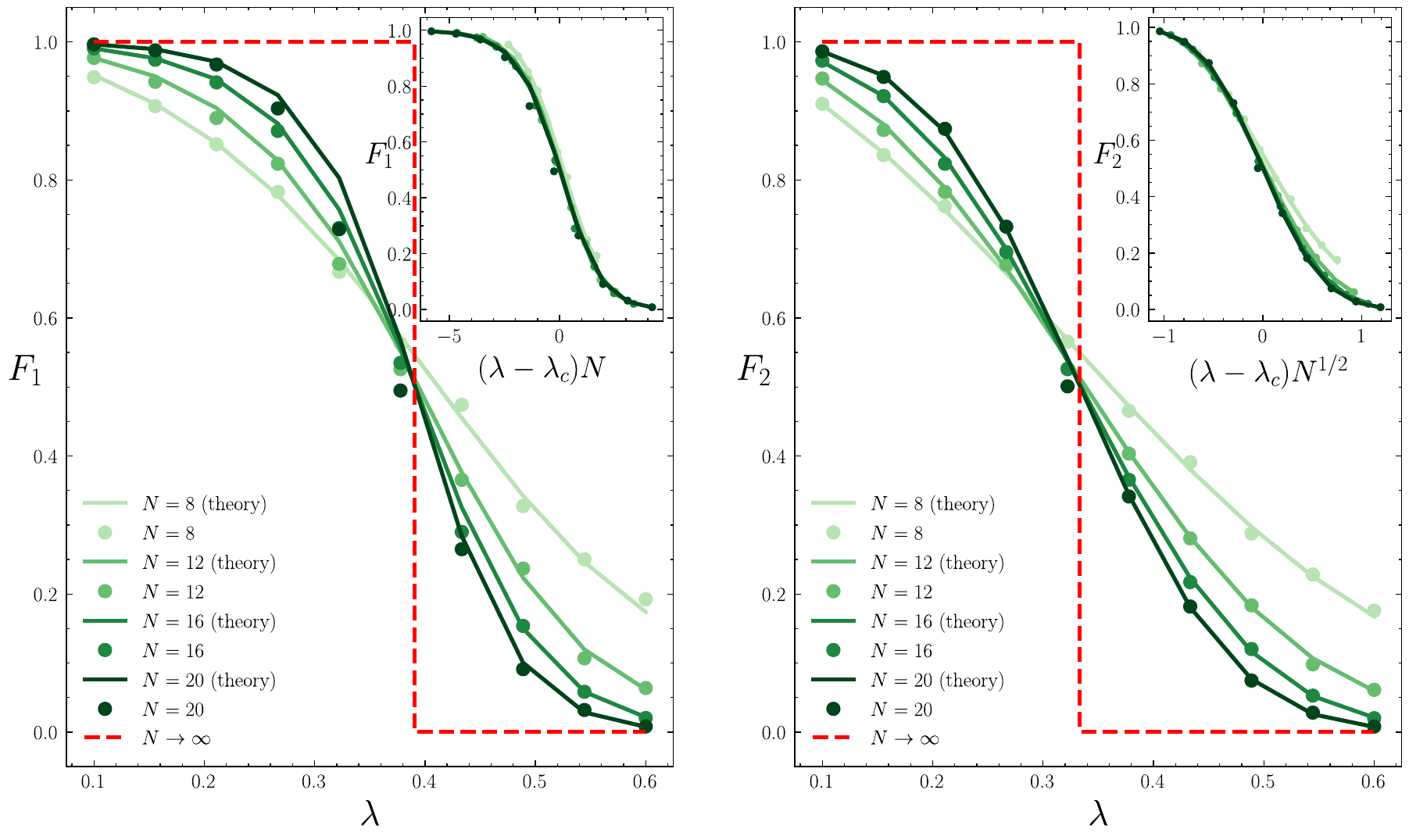}

\caption{\label{Grand Canonical Numerics} Fidelity of the $z$-only depolarization
channel with $r=\frac{1}{2}$ and $\phi=\frac{\pi}{2}$, both simulated
encoding-quenched (dots) and analytical encoding-annealed (lines)
for the grand-canonical ensemble. Left: Error-annealed. Right: Error-quenched.
Insets: A scaling plot of the same result.}
\end{figure}

Plots of the fidelities in the grand-canonical ensemble are presented
in Fig. \ref{Grand Canonical Numerics}. The self-averaging property
over the encoding space matches very well even for the low $N=8$,
and the scaling matches the expected critical exponents in both $F_{1}$
and $F_{2}$. The critical point is occurs at a different value of
$\lambda$ due to $F_{1}$ being a function of the $\phi$-independent
$\tilde{E}\left(\lambda\right)$ while $F_{2}$ depends on the $\phi$-dependent
$\bar{E}\left(\lambda,\phi\right)$. The results for $F_{1}$ match
the deterministic channel in Ref. \citep{PhysRevLett.132.140401}
as predicted.

\begin{figure}
\includegraphics[scale=0.28]{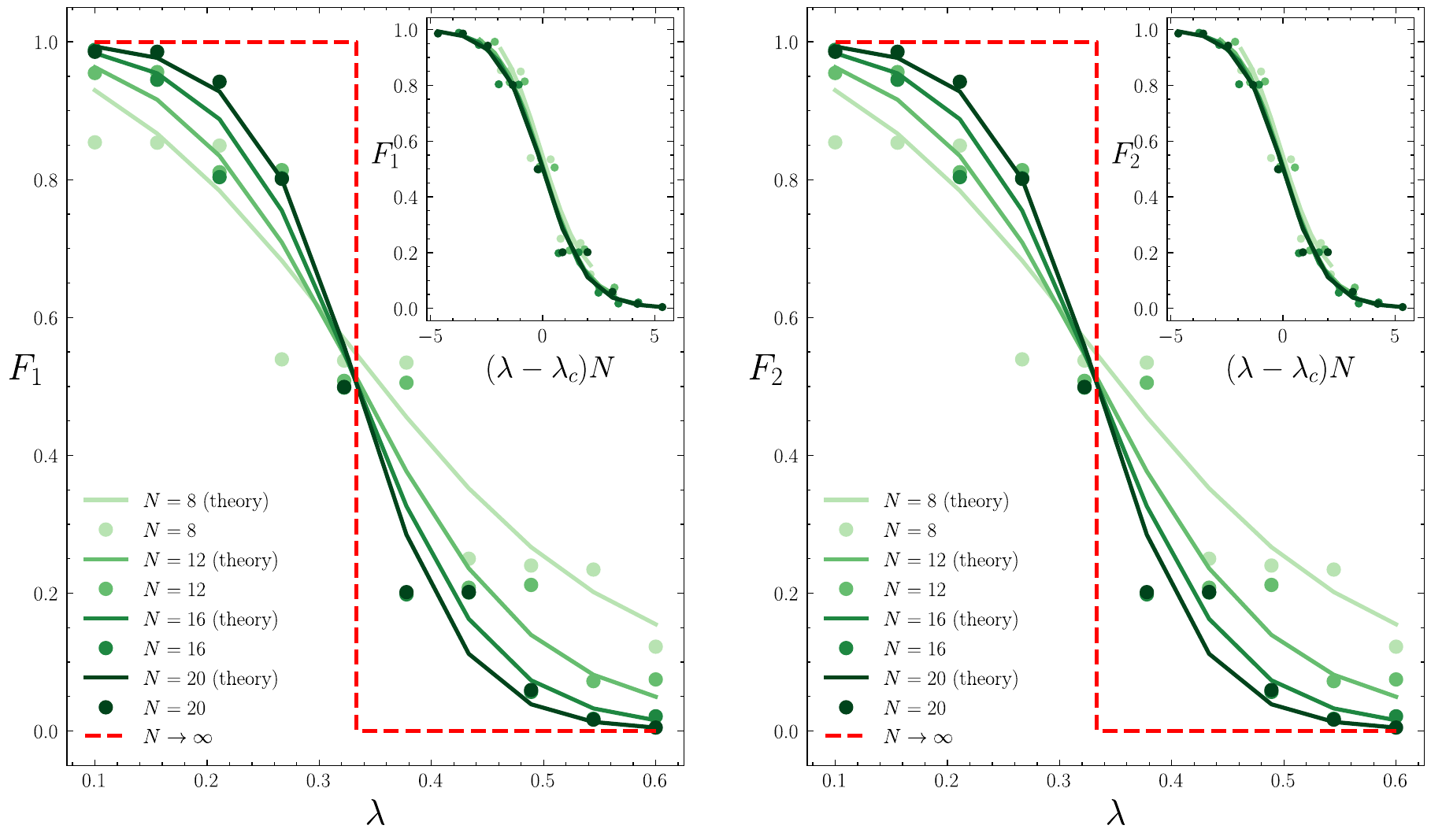}

\caption{\label{Canonical Numerics} Fidelity of the $z$-only depolarization
channel with $r=\frac{1}{2}$ and $\phi=\frac{\pi}{2}$, both simulated
encoding-quenched (dots) and analytical encoding-annealed (lines)
for the canonical ensemble. Left: Error-annealed. Right: Error-quenched.
Insets: A scaling plot of the same result.}

\end{figure}

Similar plots of the fidelities in the canonical ensemble are given
by Fig. \ref{Canonical Numerics}. The numerical results show a staircase
pattern due to the number of errors being a discrete variable, in
contrast to the continuous probabilities of the grand-canonical ensemble.
Despite this, for $N=20$, the numerical results still match the theoretical
prediction quite well far enough from the critical point. The scaling
plot shows that the numerical results cluster around the same $\nu=1$
analytical universal function, which saturates the theoretical bound.
The plots of $F_{1}$ and $F_{2}$ match each other well, as implied
by qubit-order independence.

\paragraph{Conclusions.---}

We have shown how the choice of thermodynamic ensemble can drastically
change both the critical exponent and natural parameter indpenedently,
and even completely eliminate the phase transition. This goes beyond
previous works \citep{Igloi_1998,Campa_2025,PhysRevLett.134.207401,d4wh-hqcp,1968PhRv..176..257F,Dhar_2003,Vodola_2022,PhysRevLett.79.5130,PhysRevLett.133.050403,Behringer_2005}.
It would be interesting to find other examples of the strong ensemble
dependence of critical exponents.

Moreover, we have shown how our generalized method for bounding critical
exponents using information arguments based on Ref. \citep{feldman2024informationboundsphasetransitions}
can be extended to general ensembles. Interestingly, the bound was
found to be saturated in all cases in our system; it will be interesting
to investigate the conditions under which such saturation may occur.

In almost all cases, we have also shown how the realization of a quantum
channel in terms of classical randomness can have significant effect
on the error correction behavior, an important lesson for those studying
the capabilities of such codes against different error sources.

\paragraph{Acknowledgments.---}

Our work has been supported by the Israel Science Foundation (ISF)
and the Directorate for Defense Research and Development (DDR\&D)
Grant No. 3427/21, the ISF Grant No. 1113/23, and the US-Israel Binational
Science Foundation (BSF) Grant No. 2024140.

\bibliographystyle{apsrev4-2}
\bibliography{2C__Users_idanl_Documents_Thesis_bibliography}

\end{document}


\title{Supplemental Material: Ensemble Dependence of the Critical Exponent
at a Quantum Error Correction Threshold}
\author{Idan Dror and Moshe Goldstein}
\affiliation{School of Physics and Astronomy, Tel-Aviv University, Tel Aviv 6997801,
Israel}
\begin{abstract}
In this Supplemental we fill in some technical details of our calculations.
Sec. \ref{sec:Introduction} describes our model and its use of randomness.
Sec. \ref{sec:Calculating-the-Fidelity} delves into the analytic
calculation of the exact form the different fidelity averages in the
canonical and grand-canonical ensembles. Sec. \ref{sec:Ensemble-Interpolation}
defines an interpolation between the canonical and grand-canonical
ensembles, and derives a bound on its critical exponent. Sec. \ref{sec:Summary}
provides a summary of this supplemental. Finally, appendix \ref{sec:Appendix:-Properties-of R}
contains proofs for properties of $R$.
\end{abstract}
\maketitle

\section{\label{sec:Introduction}Introduction}

\subsection{Error-Correction Toy Model}

We work with a model which is slight generalization of the encoding-decoding
circuit described in Ref. \citep{PhysRevLett.132.140401}:
\begin{itemize}
\item A quantum system starts in an arbitrary pure state of a Hilbert space
of dimension $K$, representing a logical state for computation. Without
loss of generality, we choose this state to be $\left|0\right\rangle $.
\item An ancillary quantum system initialized to $\left|0\right\rangle $
is appended to the to the logical space. The dimension of the combined
system is denoted by $D$.
\item The information of the logical part is spread over the full space
by a unitary operator, $U\in\mathcal{U}\left(D\right)$, which changes
the state of the system to be $\left|\psi\right\rangle =U\left|0,\bar{0}\right\rangle $,
where the bar denotes the ancilla.
\item The entire system is affected by some quantum channel $\mathcal{E}\left(\rho\right)=\sum_{\vec{\mu}}K_{\vec{\mu}}\rho K_{\vec{\mu}}^{\dagger}$
representing an error, with $K_{\vec{\mu}}$ being a choice of Kraus
operators for the channel and $\vec{\mu}$ is an arbitrary index.
\item The state is ``decoded'' by applying the inverse operator $U^{-1}=U^{\dagger}$.
\end{itemize}
If no error had occurred, the final state of the system would be the
same as the initial state, with the ancillary being $\left|0\right\rangle $
and the logical state intact. When an error does occur, we expect
measurements of the ancilla to act as syndromes, i.e., to yield a
non-zero result, which can be used to correct the error and retrieve
the original state, assuming the error is not too strong.

For simplicity, we only consider the no-syndrome state of the ancilla,
$\left|0\right\rangle $, and ask for the fidelity of the final logical
state with the initial state. This assumes that, for non-zero syndromes,
upon the application of error correction, one would obtain the same
fidelity of the logical state. Projecting the final state into the
no-syndrome subspace and then taking the fidelity with the original
state yields:

\begin{equation}
F=\frac{\left\langle 0\middle|\rho_{X}\middle|0\right\rangle }{\text{tr}\left(\rho_{X}\right)}=\mathbb{P}\left(0\cap\bar{0}\middle|\bar{0}\right),
\end{equation}
where $\rho_{X}=\left\langle \bar{0}\middle|U^{\dagger}\mathcal{E}\left(U\left|0,\bar{0}\right\rangle \left\langle 0,\bar{0}\right|U^{\dagger}\right)U\middle|\bar{0}\right\rangle $
is the projection of the final state into the logical subspace. 

\subsection{Probabilistic Channels}

Let us consider a convex combination of channels, that is, a set of
channels $\left\{ \mathcal{E}_{\alpha}\right\} _{\alpha}$, from which
one selects with a probability distribution $\mathbb{P}\left(\mathcal{E}_{\alpha}\right)\equiv\mathbb{P}\left(\alpha\right)\equiv p_{\alpha}$.
Given a starting state $\rho$, and applying a random channel from
that distribution, the probability of measuring $x$ (denoted as $\mathbb{P}_{\text{prob}}\left(x\right)$)
is given by:

\begin{align}
\mathbb{P}_{\text{prob}}\left(x\right) & =\sum_{\alpha}\mathbb{P}\left(\alpha\right)\mathbb{P}\left(x\middle|\alpha\right)\nonumber \\
 & =\sum_{\alpha}p_{\alpha}\left\langle x\middle|\mathcal{E}_{\alpha}\left(\rho\right)\middle|x\right\rangle \nonumber \\
 & =\left\langle x\middle|\sum_{\alpha}p_{\alpha}\mathcal{E}_{\alpha}\left(\rho\right)\middle|x\right\rangle \nonumber \\
 & \equiv\left\langle x\middle|\mathcal{E}\left(\rho\right)\middle|x\right\rangle \nonumber \\
 & =\mathbb{P}_{\text{mix}}\left(x\right),
\end{align}
where $\mathcal{E}\left(\rho\right)=\sum_{\alpha}p_{\alpha}\mathcal{E}_{\alpha}\left(\rho\right)$
is the ``mixture'' channel and the probability of measuring $x$
after applying the mixture channel to $\rho$ is $\mathbb{P}_{\text{mix}}\left(x\right)$.
We see that the two probabilities are the same, meaning that the ensemble
and the mixture channel produce indistinguishable states. However,
for each realization, the final states are distinct, $\mathcal{E}\left(\rho\right)\neq\mathcal{E}_{\alpha}\left(\rho\right)$,
and therefore can still produce different results, and specifically
different fidelities. A model with random channels is therefore more
general then a model which only has a single known channel or a random
unitary, and can describe for example randomly measuring parts of
a system.

\section{\label{sec:Calculating-the-Fidelity}Calculating the Fidelity Analytically}

\subsection{Averaging the Fidelity}

Like Ref. \citep{PhysRevLett.132.140401}, we are interested in the
average fidelity over the entire encoding space. They defined two
ways to calculate this average: the quenched,

\begin{equation}
F\left(\mathcal{E}\right)\equiv\mathbb{E}_{U}\left[F\left(\mathcal{E}\middle|U\right)\right]=\mathbb{E}_{U}\left[\frac{\mathbb{P}\left(0\cap\bar{0}\right)}{\mathbb{P}\left(\bar{0}\right)}\right],
\end{equation}
and the annealed,

\begin{equation}
\tilde{F}\left(\mathcal{E}\right)\equiv\frac{\mathbb{E}_{U}\left[\mathbb{P}\left(0\cap\bar{0}\right)\right]}{\mathbb{E}_{U}\left[\mathbb{P}\left(\bar{0}\right)\right]}.
\end{equation}
Similarly, when using a probabilistic model, we can average over the
errors in both manners. We denote these  with an index: 1 for annealed
and 2 for quenched. When the errors are treated in an annealed manner,
we can use $\mathbb{P}\left(x\right)=\sum_{\alpha}\mathbb{P}\left(\alpha\right)\mathbb{P}\left(x\middle|\alpha\right)=\mathbb{E}_{\alpha}\left[\mathbb{P}\left(x\middle|\alpha\right)\right]$
to show that the resulted average fidelity is the average fidelity
of the deterministic mixture channel:

\begin{equation}
F_{1}\left(\left\{ \left(\mathcal{E}_{\alpha},p_{\alpha}\right)\right\} _{\alpha}\right)\equiv\mathbb{E}_{U}\left[\frac{\mathbb{E}_{\alpha}\left[\mathbb{P}\left(0\cap\bar{0}|\alpha\right)\right]}{\mathbb{E}_{\alpha}\left[\mathbb{P}\left(\bar{0}|\alpha\right)\right]}\right]=\mathbb{E}_{U}\left[\frac{\mathbb{P}\left(0\cap\bar{0}\right)}{\mathbb{P}\left(\bar{0}\right)}\right]=F\left(\sum_{\alpha}p_{\alpha}\mathcal{E}_{\alpha}\right),
\end{equation}

\begin{equation}
\tilde{F}_{1}\left(\left\{ \left(\mathcal{E}_{\alpha},p_{\alpha}\right)\right\} _{\alpha}\right)\equiv\frac{\mathbb{E}_{U}\left[\mathbb{E}_{\alpha}\left[\mathbb{P}\left(0\cap\bar{0}|\alpha\right)\right]\right]}{\mathbb{E}_{U}\left[\mathbb{E}_{\alpha}\left[\mathbb{P}\left(\bar{0}|\alpha\right)\right]\right]}=\frac{\mathbb{E}_{U}\left[\mathbb{P}\left(0\cap\bar{0}\right)\right]}{\mathbb{E}_{U}\left[\mathbb{P}\left(\bar{0}\right)\right]}=\tilde{F}\left(\sum_{\alpha}p_{\alpha}\mathcal{E}_{\alpha}\right).\label{eq: F_1 is deterministic}
\end{equation}
This is not the case when considering the quenched average over the
errors:

\begin{equation}
F_{2}\left(\left\{ \left(\mathcal{E}_{\alpha},p_{\alpha}\right)\right\} _{\alpha}\right)\equiv\mathbb{E}_{\alpha}\left[F\left(\mathcal{E}_{\alpha}\right)\right],
\end{equation}

\begin{equation}
\tilde{F}_{2}\left(\left\{ \left(\mathcal{E}_{\alpha},p_{\alpha}\right)\right\} _{\alpha}\right)\equiv\mathbb{E}_{\alpha}\left[\tilde{F}\left(\mathcal{E}_{\alpha}\right)\right],
\end{equation}
for which the probabilistic model yields a different result than the
deterministic one. According to Ref. \citep{PhysRevLett.132.140401},
$F\approx\tilde{F}$ in the thermodynamic limit, and therefore we
expect $F_{1,2}\approx\tilde{F}_{1,2}$ as well.

\subsection{Calculating the Annealed Fidelity}

For a deterministic channel, calculating the average annealed fidelity
$\tilde{F}$ can be done using the replica trick from Ref. \citep{PhysRevLett.132.140401},
by calculating the average probabilities of the final states:

\begin{equation}
\mathbb{P}\left(x\cap\bar{y}\right)=\left\langle x,y\middle|A_{\mathcal{E},U}^{\left(2\right)}\middle|y,x\right\rangle ,
\end{equation}
where $A_{\mathcal{E},U}^{\left(2\right)}=\left(U^{\dagger}\right)^{\otimes2}\mathcal{K}^{\left(2\right)}U^{\otimes2}$
is the circuit applied to two copies of the system, and the error
is represented by $\mathcal{K}^{\left(2\right)}=\sum_{\vec{\mu}}K_{\vec{\mu}}\otimes K_{\vec{\mu}}^{\dagger}$,
with the $K_{\vec{\mu}}$ being the Kraus operators of the channel
$\mathcal{E}\left(\rho\right)=\sum_{\vec{\mu}}K_{\vec{\mu}}\rho K_{\vec{\mu}}^{\dagger}$.
The average probability is therefore $\mathbb{E}_{U}\left[\mathbb{P}\left(x\cap\bar{y}\right)\right]=\left\langle x,y\middle|\mathbb{E}_{U}\left[A_{\mathcal{E},U}^{\left(2\right)}\right]\middle|y,x\right\rangle \equiv\left\langle x,y\middle|A_{\mathcal{E}}^{\left(2\right)}\middle|y,x\right\rangle $.
Using the derivation in Ref. \citep{PhysRevLett.132.140401} we can
show that:

\begin{equation}
A_{\mathcal{E}}^{\left(2\right)}=\sum_{\pi,\sigma\in\mathcal{S}_{2}}W_{\pi,\sigma}\text{Tr}\left(\mathcal{K}^{\left(2\right)}T_{\sigma}\right)T_{\pi},\label{eq:A_E}
\end{equation}
where $W_{\pi,\sigma}$ are the Weingarten symbols, $\mathcal{S}_{2}$
is the permutation group of two elements, and $T_{\pi}$ being the
representations of the group on $\left(\mathbb{C}^{D}\right)^{\otimes2}$.
Hence, $\tilde{F}$ can expressed using only the two values $\text{Tr}\left(\mathcal{K}^{\left(2\right)}T_{\sigma}\right)$,
one value for each of the two generators of $\mathcal{S}_{2}$ (the
identity $T_{\left(\right)}$ and the swap $T_{\left(12\right)}$).
Since the fidelity is a ratio between a numerator and a denominator
which are both linear in these traces, it is sufficient to use a single
value, which we define as: 
\begin{equation}
R\left(\mathcal{E}\right)\equiv\frac{1}{D}\frac{\text{Tr}\left(\mathcal{K}^{\left(2\right)}T_{\left(\right)}\right)}{\text{Tr}\left(\mathcal{K}^{\left(2\right)}T_{\left(12\right)}\right)},\label{R definition}
\end{equation}
in order to determine $\tilde{F}$ completely. In fact, the denominator
in Eq. (\ref{R definition}) is constant, and hence only the numerator
needs to be recalculated for each channel and there is no division
by $0$. n the Appendix we provide proofs for this claim and for the
following important properties of $R$:
\begin{itemize}
\item $R$ is valid function the channel, meaning it does not depend on
the choice of basis or Kraus operators.
\item Bounded: $0\leq R\leq1$, with both bounds saturated. Notably, $R\left(\mathcal{E}\right)=1$
if and only if $\mathcal{E}$ is the identity channel. 
\item Linear: $R\left(\sum_{\alpha}p_{\alpha}\mathcal{E}_{\alpha}\right)=\sum_{\alpha}p_{\alpha}R\left(\mathcal{E}_{\alpha}\right)$.
\item Multiplicative: $R\left(\bigotimes_{i}\mathcal{E}_{i}\right)=\prod_{i}R\left(\mathcal{E}_{i}\right)$.
\end{itemize}
Intuitively, $R$ represents ``how much'' of of the original state
is ``retained'' after an error, which would make it negatively correlated
with a notion of error strength. The calculation of the average fidelity
will confirm this. We therefore define the error strength to be 
\begin{equation}
E\left(\mathcal{E}\right)=-\log_{D}R\left(\mathcal{E}\right)\in\left[0,\infty\right].\label{eq: E definition}
\end{equation}
Using the definition of $R$ and its property, we are able to calculate
$\tilde{F}$ and its extensions $\tilde{F}_{1,2}$ analytically in
a manner that simplifies and generalizes the results of \citep{PhysRevLett.132.140401}.

\subsection{Global Channel}

For a channel that acts globally on the entire system, we are able
to simplify $A_{\mathcal{E}}^{\left(2\right)}$ from (\ref{eq:A_E})
as

\begin{equation}
A_{\mathcal{E}}^{\left(2\right)}=\frac{1}{D^{2}-1}\left[\left(D^{2}R-1\right)T_{\left(\right)}+D\left(1-R\right)T_{\left(12\right)}\right],
\end{equation}
and use it to calculate the average probabilities in the annealed
fidelity:

\begin{equation}
\mathbb{E}_{U}\left[\mathbb{P}\left(0\cap\bar{0}\right)\right]=\frac{1}{D^{2}-1}\left[\left(D-1\right)\left(DR+1\right)\right],
\end{equation}

\begin{equation}
\mathbb{E}_{U}\left[\mathbb{P}\left(\bar{0}\right)\right]=\sum_{i=0}^{K-1}\mathbb{P}\left(i\cap\bar{0}\right)=\frac{1}{D^{2}-1}\left[DR\left(D-K\right)+KD-1\right].
\end{equation}
The annealed fidelity thus simplifies to the following form:

\begin{equation}
\tilde{F}=\frac{\left(D-1\right)\left(DR+1\right)}{DR\left(D-K\right)+KD-1}.\label{eq: base F}
\end{equation}

\subsection{Multiple Subsystems}

We now turn our attention to a system comprised of $N$ identical
sites (also known as qudits), of which $k$ are logical and the other
are ancillary. The dimension of each qudit is denoted by $d$, and
therefore dimension of the entire system is $D=d^{N}$. Similarly,
the dimension of the logical subsystem is $K=d^{k}$. The encoding
rate $r\equiv\frac{k}{N}$ is defined in the same manner as in Ref.
\citep{PhysRevLett.132.140401}. Without loss of generality, we assume
that the error channel acts on each qudit independently, i.e., $\mathcal{E}=\bigotimes_{i=1}^{N}\mathcal{E}_{i}$.
Using (\ref{eq: product}), Eq. (\ref{eq: base F}) simplifies to:

\begin{equation}
\tilde{F}=\frac{\left(d^{N}-1\right)\left(d^{N}\prod_{i}R_{i}+1\right)}{d^{N}\left(d^{N}-d^{k}\right)\prod_{i}R_{i}+d^{N+k}-1},\label{eq: F qudits}
\end{equation}
where $R_{i}=R\left(\mathcal{E}_{i}\right)$. This expression has
the same form as in Ref. \citep{PhysRevLett.132.140401} for the case
of identical channels $\mathcal{E}=\mathcal{E}_{1}^{\otimes N}$.
\Tabref{{R values}} can be used to show that the results are indeed
equal.

Note that this form is independent of the ordering of the errors.
This is expected since a reordering of the sites can be done by applying
a permutation matrix, which can always be absorbed into the average
over the encoding unitary without affecting the Haar measure.

Using the definition of $E$ in Eq. (\ref{eq: E definition}), Eq.
(\ref{eq: F qudits}) simplifies further into:

\begin{equation}
\tilde{F}=\frac{\left(d^{N}-1\right)\left(d^{N}d^{-N\bar{E}}+1\right)}{d^{N}\left(d^{N}-d^{k}\right)d^{-N\bar{E}}+d^{N+k}-1},
\end{equation}
with $\bar{E}=-\frac{1}{N}\sum_{i}\log_{d}R_{i}$ being the average
of the error strength over the sites. Taking the thermodynamic limit
$N\gg1$ while keeping $r$ constant, the fidelity goes through a
phase transition:

\begin{equation}
\tilde{F}\approx\frac{1}{1+d^{N^{1/\nu}\left(\bar{E}-E_{c}\right)}}\to\Theta\left(-\left(\bar{E}-E_{c}\right)\right),
\end{equation}
with $E_{c}\equiv1-r$ being the ``redundancy rate'' (ratio between
number of ancilla and total number of sites) and critical value, $\nu=1$
is the critical exponent, and $\Theta\left(x\right)$ is Heaviside's
step-function (with $\Theta\left(0\right)\equiv\frac{1}{2}$) . This
shows why $E$ is a natural choice for the definition of error strength
--- the phase transition occurs when the average error strength is
grater than the redundancy of the system.

Even though $\bar{E}$ is the most natural parameterization of the
fidelity, other parameterizations are possible, including, in particular,
any parameter used for defining the channel, like the probability
of a specific error channel being selected out of the ensemble. Given
a family of channels $\mathcal{E}_{\lambda}$, parameterized by some
parameter $\lambda$, the average error $\bar{E}\left(\lambda\right)$
may not cross $E_{c}$, destroying the phase transition. An example
for that is a case with a channel that has $R=0$ with a non-zero
probability, causing $\bar{E}=\infty$. In general, $\bar{E}\geq1$
is always in the ``error-preserving'' phase (zero fidelity), since
$r<1$.

\subsection{Probabilistic Model}

In the probabilistic model, the channel is chosen randomly from a
distribution. Such a distribution, when used with a separable system,
may take the form of an ensemble. We work with two ensembles:
\begin{itemize}
\item A ``grand-canonical'' ensemble, in which errors are chosen independently
from an identical and independent single-site distribution, allowing
the number of errors from each type to vary. Assuming the probability
of channel $\alpha$ to be selected for any single site is $p_{\alpha}$,
the expected number of errors of that type is $\mathbb{E}\left[N_{\alpha}\right]=p_{\alpha}N$.
\item A ``canonical'' ensemble, in which the number of errors of each
type is fixed to be $p_{\alpha}N$ to match the grand-canonical distribution,
but no variation is allowed. In such a model, the ordering of the
errors is random, and the probability for each ordering is arbitrary.
\end{itemize}
We compared both models and calculated their critical exponents. In
both cases, we assumed the channel is separable $\mathcal{E}_{\vec{\mu}}=\bigotimes_{i=1}^{N}\mathcal{E}_{\mu_{i}}$.

\subsubsection{Canonical Ensemble}

For the canonical model, the probability distribution is:

\begin{equation}
p_{\vec{\mu}}=\begin{cases}
p_{\pi} & \vec{\mu}=\pi\left(\left(\dots,\underset{p_{\alpha}N\text{ times}}{\underbrace{\dots,\alpha,\dots}},\dots\right)\right),\\
0 & \text{else},
\end{cases}
\end{equation}
where $\pi\in\mathcal{S}_{N}$ is a permutation of sites and $p_{\pi}$
is an arbitrary probability distribution over the orderings. Since
the ordering of the errors does not affect the fidelity, the expected
fidelity is the same as in the deterministic case regardless of the
ordering distribution and averaging method:

\begin{equation}
\tilde{F}_{1}\left(\left\{ \left(p_{\vec{\mu}},\mathcal{E}_{\vec{\mu}}\right)\right\} _{\vec{\mu}}\right)=\tilde{F}_{2}\left(\left\{ \left(p_{\vec{\mu}},\mathcal{E}_{\vec{\mu}}\right)\right\} _{\vec{\mu}}\right)=\tilde{F}\left(\bigotimes_{\alpha}\mathcal{E}_{\alpha}^{\otimes\left(p_{\alpha}N\right)}\right),
\end{equation}
which has the same natural parameter $\bar{E}=\frac{1}{N}\sum_{i}R_{\alpha_{i}}=\sum_{\alpha}p_{\alpha}R_{\alpha}$,
critical point $E_{c}=1-r$, and critical exponent $\nu=1$.

\subsubsection{Grand-Canonical Ensemble}

For the grand-canonical model, the probability distribution is:

\begin{equation}
p_{\vec{\mu}}=\prod_{i}p_{\mu_{i}}.
\end{equation}

\paragraph{Calculating $\tilde{F}_{1}$.}

Eq. (\ref{eq: F_1 is deterministic}) shows that $\tilde{F}_{1}$
is the same as that of a deterministic mixture channel $\sum_{\vec{\mu}}p_{\vec{\mu}}\mathcal{E}_{\vec{\mu}}$,
while Eq. (\ref{eq: base F}), shows as that it only depends on $R$
of that channel. Calculating $R$ yields:

\begin{align}
R\left(\sum_{\vec{\mu}}p_{\vec{\mu}}\mathcal{E}_{\vec{\mu}}\right) & =\sum_{\vec{\mu}}p_{\vec{\mu}}R\left(\mathcal{E}_{\vec{\mu}}\right)\nonumber \\
 & =\mathbb{E}_{\vec{\mu}}\left[R\left(\mathcal{E}_{\vec{\mu}}\right)\right]\nonumber \\
 & =\mathbb{E}_{\vec{\mu}}\left[\prod_{i}R\left(\mathcal{E}_{i}\right)\right]\nonumber \\
 & =\prod_{i}\mathbb{E}_{i}\left[R\left(\mathcal{E}_{i}\right)\right]\nonumber \\
 & =\left(\mathbb{E}_{\alpha}\left[R\left(\mathcal{E}_{\alpha}\right)\right]\right)^{N}\nonumber \\
 & =\left(\sum_{\alpha}p_{\alpha}R_{\alpha}\right)^{N},
\end{align}
where we used the independence of the sites to take the product out
of the expectation value. Therefore, $E=-N\log_{d}\left(\sum_{\alpha}p_{\alpha}R_{\alpha}\right)=NE\left(\bar{R}\right)$.
Hence, we still have $\nu=1$ and $E_{c}=1-r$, but now with a different
natural parameter $\tilde{E}=E\left(\bar{R}\right)\neq\bar{E}$. Note
that we used base $d$ for the logarithm to fit the single-site channels,
but the error strength of the global channel is base $D$.

This yields the same result (see footnote \footnote{Up to a $1-r\to r$ typo in their critical value for the coherent
channel, fixed in \citep{sierant2026theorymagicphasetransitions}}) as in Ref. \citep{PhysRevLett.132.140401}, with an identical mixture
channel duplicated across all sites $R\left(\mathcal{E}\right)=R\left(\mathcal{E}_{\text{mix}}^{\otimes N}\right)$,
which is what we would expect if the channels represented randomness.

\paragraph{Calculating $\tilde{F}_{2}$.}

To calculate $\tilde{F}_{2}=\mathbb{E}_{\vec{\mu}}\left[\tilde{F}\left(\bigotimes_{i=1}^{N}\mathcal{E}_{\mu_{i}}\right)\right]$,
we again use of the order-independence of the errors to show that
$\tilde{F}\left(\bigotimes_{i=1}^{N}\mathcal{E}_{\mu_{i}}\right)=\tilde{F}\left(n_{1},\dots,n_{m}\right)$,
where $n_{\alpha}$ is the number of errors of type $\alpha$, and
hence $\mathbb{E}_{\vec{n}}\left[\tilde{F}\left(\vec{n}\right)\right]$.
Moving forward, we will assume without loss of generality that the
values of $R_{\alpha}$ are unique. Note that for $m=1$, there is
no randomness and the setup is therefore equivalent to the previous
cases. Therefore, the following calculation assumes $m>1$.

In the thermodynamic limit, the sum can be approximated by an integral:
\begin{align}
\sum_{\left\{ n_{\alpha}\right\} :\sum_{\alpha}n_{\alpha}=N} & =\sum_{\left\{ n_{\alpha}\right\} }\delta_{\sum_{\alpha}n_{\alpha},N}\nonumber \\
 & \to\int_{0}^{1}N^{m}d^{m}x\delta\left(N\left(\sum_{\alpha}x_{\alpha}-1\right)\right)\nonumber \\
 & =\int_{0}^{1}N^{m-1}d^{m}x\delta\left(\sum_{\alpha}x_{\alpha}-1\right),
\end{align}
where $x_{\alpha}\equiv\frac{n_{\alpha}}{N}$. Noting that the distribution
of $\vec{n}$ is multinomial, $\mathbb{P}\left(\vec{n}\right)=N!\prod_{\alpha}\frac{p_{\alpha}^{n_{\alpha}}}{n_{\alpha}!}$,
it is useful to invoke Stirling's approximation, $n!\approx\sqrt{2\pi n}\left(\frac{n}{e}\right)^{n}$,
to find:

\begin{align}
\frac{N!}{\prod_{\alpha}n_{\alpha}!} & \approx\frac{\sqrt{2\pi N}\left(\frac{N}{e}\right)^{N}}{\sqrt{2\pi}^{m}\prod_{\alpha}\sqrt{n_{\alpha}}\left(\frac{n_{\alpha}}{e}\right)^{n_{\alpha}}}\nonumber \\
 & =\frac{\sqrt{N}}{\sqrt{2\pi}^{m-1}}\frac{N^{N}}{\prod_{i}n_{\alpha}^{n_{\alpha}+\frac{1}{2}}}\nonumber \\
 & =\frac{\sqrt{N}}{\sqrt{2\pi}^{m-1}}\frac{\left(\frac{N}{\prod_{\alpha}\left(Nx_{\alpha}\right)^{x_{\alpha}}}\right)^{N}}{\prod_{\alpha}\sqrt{Nx_{\alpha}}}\nonumber \\
 & =\frac{1}{\sqrt{2\pi N}^{m-1}}\frac{\left(\prod_{\alpha}x_{\alpha}^{x_{\alpha}}\right)}{\sqrt{\prod_{\alpha}x_{\alpha}}}^{-N}.
\end{align}
Note that we used $\sum_{\alpha}n_{\alpha}=N$, which is enforced
by the $\delta$-function. Together, we get the following form for
$\tilde{F}_{2}$:

\begin{align}
\tilde{F}_{2} & \approx\sqrt{\frac{N}{2\pi}}^{m-1}\int_{0}^{1}d^{m}x\delta\left(\sum_{\alpha}x_{\alpha}-1\right)\exp\left(-Nf\left(\vec{x}\right)\right),
\end{align}
where: 
\begin{equation}
f\left(\vec{x}\right)\equiv\sum_{\alpha}x_{\alpha}\ln\left(\frac{x_{\alpha}}{p_{\alpha}}\right)+\frac{1}{N}\ln\sqrt{\prod x_{\alpha}}+\frac{1}{N}\ln\left[1+\left(\frac{d^{r-1}}{\prod_{\alpha}R_{\alpha}^{x_{\alpha}}}\right)^{N}\right].
\end{equation}
The first term in $f\left(\vec{x}\right)$ is the relative entropy,
$D_{\text{KL}}\left(\vec{x}\parallel\vec{p}\right)$, and is therefore
non-negative. The last term is a logarithm of a value which is greater
than one, which is also non-negative. The middle term is negative,
but negligible, since it is $O\left(\frac{1}{N}\right)$ while the
other terms are $O\left(1\right)$, hence any positive positive contribution
from the first and last term are large enough to keep the entire expression
positive, for some appropriately large $N$. Therefore, for the exponent
to contribute in the thermodynamic limit, the first and last terms
must both be zero. Let us consider them separately:
\begin{itemize}
\item The first term vanishes only for $x_{\alpha}=p_{\alpha}$ when enforcing
$\sum_{\alpha}x_{\alpha}=1$.
\item The second term vanishes only for $E\left(\vec{x}\right)\equiv-\sum_{\alpha}x_{\alpha}\log_{d}R_{\alpha}<E_{c}$.
Under this condition, the derivatives of this term are also negligible.
\end{itemize}
Using the saddle-point method, the fidelity becomes:

\begin{equation}
\tilde{F}_{2}\approx\frac{\sqrt{\frac{N}{2\pi}}^{m-1}}{\sqrt{\prod_{\alpha}p_{\alpha}}}\int_{-\infty}^{\infty}d^{m}x\delta\left(\sum_{\alpha}x_{\alpha}-1\right)\Theta\left(E_{c}-\vec{E}\cdot\vec{x}\right)e^{-N\sum_{\alpha}\frac{\left(x_{\alpha}-p_{\alpha}\right)^{2}}{2p_{\alpha}}},
\end{equation}
where $E_{\alpha}\equiv-\log_{d}R_{\alpha}$. Shifting $x_{\alpha}-p_{\alpha}\equiv y_{\alpha}$
yields:

\begin{equation}
\tilde{F}_{2}\approx\frac{\sqrt{\frac{N}{2\pi}}^{m-1}}{\sqrt{\prod_{\alpha}p_{\alpha}}}\int_{-\infty}^{\infty}d^{m}y\delta\left(\sum_{\alpha}y_{\alpha}\right)\Theta\left(E_{c}-\bar{E}-\vec{E}\cdot\vec{y}\right)e^{-N\sum_{\alpha}\frac{y_{\alpha}^{2}}{2p_{\alpha}}},
\end{equation}
which, after rescaling $N\frac{y_{\alpha}^{2}}{2p_{\alpha}}\equiv u_{\alpha}^{2}$,
becomes:

\begin{equation}
\tilde{F}_{2}\approx\frac{1}{\sqrt{\pi^{m-1}}}\int_{-\infty}^{\infty}d^{m}u\delta\left(\hat{p}\cdot\vec{u}\right)\Theta\left(E_{c}-\bar{E}-\sqrt{\frac{2}{N}}\vec{P}\cdot\vec{u}\right)e^{-u^{2}},
\end{equation}
where $\hat{p}_{\alpha}\equiv\sqrt{p_{\alpha}}$ and $P_{\alpha}=E_{\alpha}\sqrt{p_{\alpha}}$.
Rotating into new coordinate system $\vec{v}$ in where the $m$-th
direction is parallel to $\hat{p}$, we split $\vec{P}$ into perpendicular
and parallel components to $\hat{p}$, $\vec{a}\equiv\vec{a}_{\perp}+\left(\hat{a}\cdot\hat{p}\right)\hat{p}$,
and integrate over $v_{m}$ to obtain:

\begin{equation}
\tilde{F}_{2}\approx\frac{1}{\sqrt{\pi^{m-1}}}\int_{-\infty}^{\infty}d^{m-1}v\Theta\left(E_{c}-\bar{E}-\sqrt{\frac{2}{N}}\vec{P}_{\perp}\cdot\vec{v}\right)e^{-v^{2}},
\end{equation}
in which all vectors have been projected into the $m-1$ dimensional
perpendicular subspace. Splitting the projected $\vec{v}$ with respect
to the projected $\vec{P}_{\perp}$ as $\vec{v}=\vec{v}_{\perp}+v_{\parallel}\hat{P}_{\perp}$
yields:

\begin{align}
\tilde{F}_{2} & \approx\frac{1}{\sqrt{\pi^{m-1}}}\int_{-\infty}^{\infty}d^{m-1}v\Theta\left(E_{c}-\bar{E}-\sqrt{\frac{2}{N}}\left|\vec{P}_{\perp}\right|v_{\parallel}\right)e^{-v^{2}}\nonumber \\
 & =\frac{1}{\sqrt{\pi}}\int_{-\infty}^{\infty}dv_{\parallel}\Theta\left(E_{c}-\bar{E}-\sqrt{\frac{2}{N}}\left|\vec{P}_{\perp}\right|v_{\parallel}\right)e^{-v_{\parallel}^{2}}\nonumber \\
 & =\frac{1}{\sqrt{\pi}}\int_{-\infty}^{\sqrt{\frac{N}{2P_{\perp}^{2}}}\left(E_{c}-\bar{E}\right)}dv_{\parallel}e^{-v_{\parallel}^{2}}\nonumber \\
 & =\frac{1}{2}\left(\text{erf}\left(-\sqrt{\frac{N}{2P_{\perp}^{2}}}\left(\bar{E}-E_{c}\right)\right)+1\right).
\end{align}
We can also write $P_{\perp}^{2}$ explicitly as:

\begin{align}
\vec{P_{\perp}}^{2} & =\vec{P}^{2}-\left(\vec{P}\cdot\hat{p}\right)^{2}\nonumber \\
 & =\sum_{\alpha}p_{\alpha}\left(\log_{d}R_{\alpha}\right)^{2}-\left(\sum_{\alpha}p_{\alpha}\log_{d}R_{\alpha}\right)^{2}\nonumber \\
 & =\text{Var}\left(E\right),
\end{align}
which does not depend $N$, and therefore shows that $\nu=2$ and
that the critical point is at $E_{c}$ for the natural parameter $\bar{E}$.
Note that by demanding the $R_{\alpha}$ to be unique, and $m>1$
we get $\vec{P_{\perp}}^{2}=\text{Var}\left(E\right)>0$. The final
form of $\tilde{F}_{2}$ is then given entirely in terms of the distribution
of the error strength $E$:

\begin{equation}
\tilde{F}_{2}\approx\frac{1}{2}\left(\text{erf}\left(-\sqrt{\frac{N}{2\text{Var}\left(E\right)}}\left(\bar{E}-E_{c}\right)\right)+1\right)\to\Theta\left(-\left(\bar{E}-E_{c}\right)\right).
\end{equation}

More generally, since the errors are independently and identically
distributed, then in the thermodynamic limit the statistical average
of the error strength over the qubits $E_{\text{avg}}$ is normally
distributed $E_{\text{avg}}\sim\mathcal{N}\left(\bar{E},\sqrt{\frac{\text{Var}\left(E\right)}{N}}\right)$.
Because the fidelity of any specific selection of errors depend only
on $E_{\text{avg}}$, the mean fidelity can be estimated by the Gaussian
integral:

\begin{align*}
\tilde{F_{2}} & =\int\mathbb{P}\left(E_{\text{avg}}\right)F\left(E_{\text{avg}}\right)dE_{\text{avg}}\\
 & \approx\int\sqrt{\frac{N}{2\pi\text{Var}\left(E\right)}}e^{-\frac{N\left(E_{\text{avg}}-\bar{E}\right)^{2}}{2\text{Var}\left(E\right)}}f\left(N^{1/\nu_{1}}\cdot\left(E_{\text{avg}}-E_{c}\right)\right)dE_{\text{avg}}\\
 & =\frac{1}{\sqrt{\pi}}\int e^{-\left(x-\sqrt{\frac{N}{2\text{Var}\left(E\right)}}\bar{E}\right)^{2}}f\left(N^{1/\nu_{1}}\left(\sqrt{\frac{2\text{Var}\left(E\right)}{N}}x-E_{c}\right)\right)dx\\
 & \approx\frac{1}{\sqrt{\pi}}\int e^{-\left(x-\sqrt{\frac{N}{2\text{Var}\left(E\right)}}\bar{E}\right)^{2}}\Theta\left(E_{c}-\sqrt{\frac{2\text{Var}\left(E\right)}{N}}x\right)dx\\
 & =\frac{1}{\sqrt{\pi}}\int_{-\infty}^{\sqrt{\frac{N}{2\text{Var}\left(E\right)}}E_{c}}e^{-\left(x-\sqrt{\frac{N}{2\text{Var}\left(E\right)}}\bar{E}\right)^{2}}dx\\
 & =\frac{1}{2}\left(\text{erf}\left(-\sqrt{\frac{N}{2\text{Var}\left(E\right)}}\left(\bar{E}-E_{c}\right)\right)+1\right),
\end{align*}
Which yields the same result regardless of the exact form of $F$
near the critical point, assuming $\nu_{1}<2$, which is satisfied
for the fidelity $\nu_{1}=1$.

\section{\label{sec:Ensemble-Interpolation}Ensemble Interpolation}

Our analytical calculations and simulations are focused on the most
used ensembles --- the canonical and grand-canonical. Both ensembles
can be viewed as limiting cases of a multivariable hypergeometric
distribution \citep{johnson1997hypergeometric}. Suppose that there
is a ``reservoir'' of errors, its total size is $N_{\text{bath}}$,
distributed among error types $\alpha$, such that each is realized
exactly $p_{\alpha}N_{\text{bath}}$ times. To construct a circuit,
we take $N\leq N_{\text{bath}}$ errors out of the reservoir, placing
one at each qubit. The number of errors still in the reservoir is
denoted as $N_{\text{hidden}}=N_{\text{bath}}-N$. For $N_{\text{bath}}\gg N$,
the distribution is effectively unchanged by the sampling, and we
have the usual way of describing the grand-canonical ensemble. Taking
instead $N_{\text{bath}}=N$, then the same number of each type of
error is always sampled, which is how we defined the canonical ensemble.

By the known properties of the multivariate hypergeometric distribution,
the variance of each type of error is given by

\begin{equation}
\text{Var}\left(N_{\alpha}\right)=N\frac{N_{\text{hidden}}}{N_{\text{bath}}-1}p_{\alpha}\left(1-p_{\alpha}\right).
\end{equation}
Taking $N_{\text{hidden}}\sim N^{a}$ for some $a\geq0$ defines our
ensemble. The variance thus scales as

\begin{equation}
\text{Var}\left(N_{\alpha}\right)\sim N^{\min\left(a,1\right)}.
\end{equation}
Suppose that we are estimating $p_{\alpha}$ as a proxy for $\lambda$.
Using Chebyshev's inequality, we obtain

\begin{equation}
\mathbb{P}\left(\left|\frac{N_{\alpha}}{N}-p_{\alpha}\right|\geq\delta\right)\leq\frac{\text{Var}\left(\frac{N_{\alpha}}{N}\right)}{\delta^{2}}\sim N^{\min\left(a,1\right)-2}\delta^{-2},
\end{equation}
hence, for $O\left(1\right)$ accuracy, we must have 
\begin{equation}
N_{\text{opt}}\apprge\delta^{-\frac{2}{2-\min\left(a,1\right)}}.
\end{equation}
Following Ref. \citep{feldman2024informationboundsphasetransitions},
to extract the sign of $\delta$ when its magnitude is known requires
$N_{\text{PT}}=N\sim\delta^{-\nu}$ samples. Since using the phase
transition simulation cannot be more efficient than working directly
with the samples from the distribution, we must have $N_{\text{PT}}\apprge N_{\text{opt}}$,
from which we can derive the bound

\begin{equation}
\nu\geq\frac{2}{2-\min\left(a,1\right)}.\label{eq:genrelized exponent}
\end{equation}
For $a=0$ (finitely many error left out), we get $\nu\geq1$, the
same value as the canonical ensemble. For $a\geq1$, we get $\nu\geq2$,
the same value as the grand-canonical distribution, and therefore
quantifies how large must a reservoir be in order to be considered
effectively grand-canonical. The intermediate interval $0<a<1$ provides
a continuum of $\nu$ bounds interpolating between both extremes. 

\section{\label{sec:Summary}Summary}

These calculations show how the selection of ensembles, and whether
classical randomness is treated explicitly or absorbed into a deterministic
channel definition, can both affect the critical exponent and the
role of the randomness in the system. The results are summarized in
\tabref{summary_table}. For $\tilde{F}_{1}$ and the canonical ensemble,
the randomness is artificial, as there is always an equivalent deterministic
channel, while for $\tilde{F}_{2}$ in the grand-canonical ensemble,
randomness is inherent, which also changes the critical exponent from
$\nu=1$ to $\nu=2$. For $\tilde{F}_{1}$, even though the critical
exponent is the same in both ensembles, the natural parameter is ensemble
dependent. When taking a single error type $m=1$, we get $\bar{E}=\tilde{E}$
and all cases coincide with $\nu=1$. In all the cases, the limiting
step-function behavior in general and the critical value $E_{c}=1-r$
are the same.

\begin{table}[H]
\centering
\begin{tabular}{|c|c|c|}
\hline 
Ensemble \& Averaging &
Critical Exponent &
Natural Parameter\tabularnewline
\hline 
\hline 
Canonical &
$1$ &
$\bar{E}$\tabularnewline
\hline 
Grand-Canonical $\tilde{F}_{1}$ &
$1$ &
$\tilde{E}=E\left(\bar{R}\right)$\tabularnewline
\hline 
Grand-Canonical $\tilde{F}_{2}$ &
$2$ &
$\bar{E}$\tabularnewline
\hline 
\end{tabular}\caption{Comparison of the critical exponent and natural parameter for the
different ensembles and averaging methods for $m>1$.\label{tab:summary_table}}
\end{table}

\appendix

\section*{\label{sec:Appendix:-Properties-of R}Appendix: Properties of $R$}

Throughout the text we used numerous properties of $R$. These properties
are proven in this appendix.

\subsection{Independence of $T_{\left(12\right)}$}

We have claimed that the denominator of $R$, that is, $S\equiv\text{Tr}\left(\mathcal{K}^{\left(2\right)}T_{\left(12\right)}\right)$,
is in fact constant. This is straightforward from its definition:

\begin{align}
S & =\text{Tr}\left(\mathcal{K}^{\left(2\right)}T_{\left(12\right)}\right)\nonumber \\
 & =\sum_{i,j}\left\langle i,j\middle|\mathcal{K}^{\left(2\right)}T_{\left(12\right)}\middle|i,j\right\rangle \nonumber \\
 & =\sum_{i,j}\left\langle i,j\middle|\mathcal{K}^{\left(2\right)}\middle|j,i\right\rangle \nonumber \\
 & =\sum_{i,j}\left\langle i,j\middle|\sum_{\vec{\mu}}K_{\vec{\mu}}\otimes K_{\vec{\mu}}^{\dagger}\middle|j,i\right\rangle \nonumber \\
 & =\sum_{\vec{\mu}}\sum_{i,j}K_{\vec{\mu}}^{ij}K_{\vec{\mu}}^{\dagger ji}\nonumber \\
 & =\sum_{\vec{\mu}}\text{Tr}\left(K_{\vec{\mu}}K_{\vec{\mu}}^{\dagger}\right)\nonumber \\
 & =\text{Tr}\left(\mathcal{E}\left(I_{D}\right)\right)\nonumber \\
 & =\text{Tr}\left(I_{D}\right)\nonumber \\
 & =D,
\end{align}
where the last few steps stem from the channel's trace preserving
property. We therefore have $R\left(\mathcal{E}\right)=\frac{1}{D^{2}}\text{Tr}\left(\mathcal{K}^{\left(2\right)}\right)$,
which is the definition used in the rest of this supplemental and
the main text.

\subsection{$R$ is a Property of the Channel}

One might worry that we cannot uniquely define the value of $R$ for
any given channel, as its value might depend on the choice of basis
or Kraus operators. In this section we will show that this is not
the case, and the value of $R$ is truly a property of the the channel.

\subsubsection{Basis Independence}

Showing that $R$ is basis independent is a also straightforward from
its definition: 
\begin{equation}
R\left(\mathcal{E}\right)=\frac{1}{D^{2}}\text{Tr}\left(\mathcal{K}^{\left(2\right)}\right)=\frac{1}{D^{2}}\sum_{\vec{\mu}}\left|\text{Tr}\left(K_{\vec{\mu}}\right)\right|^{2},
\end{equation}
which is manifestly basis independent.

\subsubsection{Representation Independence}

To show that $R$ is also not dependent on the choice of Kraus operators,
we choose an arbitrary orthonormal basis $\left\{ \left|i\right\rangle \right\} _{i}$
of the Hilbert space. $R$ takes the following form:

\begin{align}
R\left(\mathcal{E}\right) & =\frac{1}{D^{2}}\sum_{\vec{\mu}}\left|\text{Tr}\left(K_{\vec{\mu}}\right)\right|^{2}\nonumber \\
 & =\frac{1}{D^{2}}\sum_{\vec{\mu}}\left|\sum_{i}\left\langle i\middle|K_{\vec{\mu}}\middle|i\right\rangle \right|^{2}\nonumber \\
 & =\frac{1}{D^{2}}\sum_{\vec{\mu}}\sum_{i,j}\left\langle i\middle|K_{\vec{\mu}}\middle|i\right\rangle \left\langle j\middle|K_{\vec{\mu}}^{\dagger}\middle|j\right\rangle \nonumber \\
 & =\frac{1}{D^{2}}\sum_{i,j}\left\langle i\right|\sum_{\vec{\mu}}\left(K_{\vec{\mu}}\left|i\right\rangle \left\langle j\right|K_{\vec{\mu}}^{\dagger}\right)\left|j\right\rangle \nonumber \\
 & =\frac{1}{D^{2}}\sum_{i,j}\left\langle i\right|\mathcal{E}\left(\left|i\right\rangle \left\langle j\right|\right)\left|j\right\rangle ,\label{kraus independent R}
\end{align}
which is manifestly Kraus representation independent.

\subsection{Bounds}

It is useful to know what values of $R$ are possible and which channels
produce these value.

\subsubsection{Lower bound.}

$R$ is the sum of non-negative values, and therefore $R\geq0$. This
bound is saturated by using traceless Kraus operators.

\subsubsection{Upper bound.}

Denoting $\lambda_{\vec{\mu}}^{\left(i\right)}$ as the eigenvalues
of the corresponding Kraus, $R$ takes the form

\begin{equation}
R=\frac{1}{D^{2}}\sum_{\vec{\mu}}\left|\text{Tr}\left(K_{\vec{\mu}}\right)\right|^{2}=\frac{1}{D^{2}}\sum_{\vec{\mu}}\left|\sum_{i}\lambda_{\vec{\mu}}^{\left(i\right)}\right|^{2}.
\end{equation}
Using the Lagrangian, $L=R-\left(S-D\right)\xi$, with $\xi$ being
a Lagrange multiplier:

\begin{equation}
L=\frac{1}{D^{2}}\sum_{\vec{\mu}}\left|\sum_{i}\lambda_{\vec{\mu}}^{\left(i\right)}\right|^{2}-\left(\sum_{\vec{\mu}}\sum_{i}\left|\lambda_{\vec{\mu}}^{\left(i\right)}\right|^{2}-D\right)\xi,
\end{equation}
we get the following extremum criteria:

\begin{equation}
0=\frac{\partial L}{\partial\lambda_{\vec{\mu}}^{\left(i\right)*}}=\frac{1}{D^{2}}\sum_{j}\lambda_{\vec{\mu}}^{\left(j\right)}-\xi\lambda_{\vec{\mu}}^{\left(i\right)}=\frac{\text{Tr}\left(K_{\mu}\right)}{D^{2}}-\xi\lambda_{\vec{\mu}}^{\left(i\right)}.
\end{equation}
Hence, all eigenvalues are equal, $\lambda_{\vec{\mu}}^{\left(i\right)}=\frac{\text{Tr}\left(K_{\mu}\right)}{\xi D^{2}}$,
and we deduce that $K_{\vec{\mu}}\propto I_{D}$. Denoting $K_{\vec{\mu}}=\lambda_{\vec{\mu}}I_{D}$,
we can show that $\mathcal{E}$ must be the identity channel $\mathcal{E}\left(\rho\right)=\sum_{\vec{\mu}}\left|\lambda_{\vec{\mu}}\right|^{2}\rho=\rho$.
The $R$ value of the identity channel is $R=\frac{1}{D^{2}}\sum_{\vec{\mu}}\left|\lambda_{\vec{\mu}}D\right|^{2}=1$,
and therefore $R\leq1$, and the only channel that saturate it is
the identity channel.

\subsection{Combining Channels}

There are many ways to combine channels, and for some of these the
$R$ value of the combination can be calculated from the $R$ value
of its constituents.

\subsubsection{Linear}

Given a mixture channel, $\mathcal{E}\left(\rho\right)=\sum_{\alpha}p_{\alpha}\mathcal{E}_{\alpha}\left(\rho\right)$,
we show that $R\left(\mathcal{E}\right)=\sum_{\alpha}p_{\alpha}R\left(\mathcal{E}_{\alpha}\right)$,
i.e., that $R$ is a linear. Using Eq. (\ref{kraus independent R}):

\begin{align}
R & =\frac{1}{D^{2}}\sum_{i,j}\left\langle i\right|\mathcal{E}\left(\left|i\right\rangle \left\langle j\right|\right)\left|j\right\rangle \nonumber \\
 & =\frac{1}{D^{2}}\sum_{i,j}\left\langle i\right|\sum_{\alpha}p_{\alpha}\mathcal{E}_{\alpha}\left(\left|i\right\rangle \left\langle j\right|\right)\left|j\right\rangle \nonumber \\
 & =\sum_{\alpha}p_{\alpha}\frac{1}{D^{2}}\sum_{i,j}\left\langle i\right|\mathcal{E}_{\alpha}\left(\left|i\right\rangle \left\langle j\right|\right)\left|j\right\rangle \nonumber \\
 & =\sum_{\alpha}p_{\alpha}R\left(\mathcal{E}_{\alpha}\right).
\end{align}

\subsubsection{Tensor Product}

Given a channel which operates on multiple subsystems independently,
$\mathcal{E}\left(\bigotimes_{s}\rho_{s}\right)=\bigotimes_{s}\mathcal{E}_{s}\left(\rho_{s}\right)$,
we show that $R\left(\mathcal{E}\right)=\prod_{s}R\left(\mathcal{E}_{s}\right)$:

\begin{align}
R & =\frac{1}{D^{2}}\sum_{i,j}\left\langle i\right|\mathcal{E}\left(\left|i\right\rangle \left\langle j\right|\right)\left|j\right\rangle \nonumber \\
 & =\frac{1}{D^{2}}\sum_{i,j}\left(\mathcal{E}_{s}\left\langle i_{s}\right|\right)\left(\bigotimes_{s}\mathcal{E}_{s}\left(\left|i_{s}\right\rangle \left\langle j_{s}\right|\right)\right)\left(\left|j_{s}\right\rangle \right)\nonumber \\
 & =\frac{1}{\prod_{s}D_{s}^{2}}\bigotimes_{s}\sum_{i_{s},j_{s}}\left\langle i_{s}\right|\mathcal{E}_{s}\left(\left|i_{s}\right\rangle \left\langle j_{s}\right|\right)\left|j_{s}\right\rangle \nonumber \\
 & =\prod_{s}R\left(\mathcal{E}_{s}\right).\label{eq: product}
\end{align}

\subsection{Computed Values}

In \tabref{{R values}} we present some values of $R$ as a reference
and for comparison with Ref. \citep{PhysRevLett.132.140401}.

\begin{table}[h]
\begin{tabular}{|c|c|c|c|}
\hline 
Name &
$\mathcal{E}\left(\rho\right)$ &
$R$ &
Notes\tabularnewline
\hline 
\hline 
Unitary &
$e^{i\frac{\alpha}{2}T}\rho e^{-i\frac{\alpha}{2}T}$ &
$\cos^{2}\left(\frac{\alpha}{2}\right)$ &
$T\in\mathfrak{su}\left(D\right),T^{2}=I_{D}$\tabularnewline
\hline 
Depolarization &
$\left(1-\lambda\right)\rho+\lambda\frac{I_{D}}{D}$ &
$1-\left(1-\frac{1}{D^{2}}\right)\lambda$ &
$R=\left(1-\lambda\right)R_{I}+\lambda R_{\text{reset}}$\tabularnewline
\hline 
Reset &
$\text{Tr}\left(\rho\right)\rho^{\prime}$ &
$\frac{1}{D^{2}}$ &
$\text{Tr}\rho^{\prime}=1$\tabularnewline
\hline 
Measurement &
$\sum_{i}\rho_{ii}\left|i\right\rangle \left\langle i\right|$ &
$\frac{1}{D}$ &
$\left\langle i|j\right\rangle =\delta_{ij}$\tabularnewline
\hline 
\end{tabular}

\caption{$R$ values for common channels\label{tab:R values}.}
\end{table}

\bibliographystyle{apsrev4-2}
\bibliography{2C__Users_idanl_Documents_Thesis_bibliography}